\documentclass[12pt]{article}
\begin{document}

\title{\ Superconformal anomaly free models in $D=4$}
\author{Michael Hewitt \\
Science Research Unit, Canterbury Christ Church University College,\\
North Holmes Road, Canterbury, CT1 1QU, U.K. \\ email:
mike@cant.ac.uk,
 tel:+44(0)1227-767700.}
\date{10 December 2003}
\maketitle
\begin{abstract}

A family of modified $(0,1)$ heterotic string models in $D=4$ is
constructed in which the strings incorporate $R$ flux tubes which
may in special cases support a local spacetime superconformal
symmetry consistent with quantum mechanics. There is an intrinsic
Goldstino multiplet, so that supersymmetry breaking can be driven
by any process that generates a non-zero value for the
superpotential. The superconformal anomaly freedom of these models
may be related to a naturally vanishing $\Lambda$. The $R$ charge
of such models may also play a role in producing a
generation-ordered fermion spectrum. \

\


\end{abstract}

\section{Introduction}
Our aim in this paper is to argue that strictly 4 dimensional
(0,1) heterotic string models have special features that may
enable them to provide the propagating states for models with a
local space-time superconformal invariance. By strictly 4
dimensional models we mean the sub-family of those studied in
[1,2,3] which have a field independent value of $g$, the bare
dimensionless 4 dimensional coupling constant. When the effective
level zero field theory is expressed in string tension units, the
dilaton is replaced by a residual scalar field $\psi$, with $g$
independent of $\psi$. With suitable twisted boundary conditions,
all breathing modes which have an effect on $g$ may be eliminated
from the level zero spectrum allowing such models to be
constructed. We may now introduce a zero norm chiral scalar
superfield $\chi$ (i.e. $\chi$ has no kinetic term) to compensate
the string tension in an attempt to make a model with local
superconformal invariance. This will be a symmetry of the lifted
metric
\begin{equation}\label{gh}
    \hat{g}_{\mu \nu} = \mathrm{e}^{\mathrm{Re}(\chi)}g_{\mu \nu}
\end{equation}
where conformal rescalings of $\hat{g}_{\mu \nu}$ can be absorbed
into the compensator $\chi$ with $g_{\mu \nu}$ the metric relative
to the string tension. The phase of $\chi$ compensates the phase
of mass terms for the massive Ramond sector states, respecting $R$
invariance, contributing a trivial volume factor to the path
integral. The spinor components of $\chi$ may be gauged away using
$S$ type supersymmetry transformations, as usual for a conformal
compensator [4]. The motivation here is to study supersymmetry
breaking in the context of superconformal symmetry, as proves
useful for conventional supergravity [4]. There is, in general, an
obstruction to the validity of superconformal symmetry in
conventional field theory in the form of ABJ anomalies [5,6] for
$R$ transformations. By identifying $R$ with a suitable `handle'
charge in the 2 fermion Lie subalgebra of the Ramond Clifford
algebra, it may be possible to arrange for all ABJ anomalies
involving $R$ to cancel. This extra requirement would give a
powerful constraint on the string models that could consistently
be used in this way. A consideration of the non-perturbative
defect around a string leads to a proposal that the strings should
incorporate intrinsic $R$ flux tubes. This suggests that the
strings may be topological features associated with a discrete
fibre bundle. A consequence of these intrinsic $R$ flux tubes is
that the supermultiplet containing $\psi$ becomes the Goldstone
multiplet for local supersymmetry breaking, which will now occur
automatically, provided that the superpotential $\mathcal{G}$
takes a non-zero value. If the v.e.v of $\mathcal{G}$ is generated
by processes well below the string tension scale, this would lead
to a stable realistic supersymmetry breaking mechanism.  It may be
possible to establish a theorem giving zero cosmological constant
in the form of the somewhat stronger statement that the global
superconformal invariance of the global (vacuum) state is anomaly
free even though it is spontaneously broken. A characteristic
feature of this realisation of superconformal invariance is that
the antisymmetric tensor counterpart of the graviton is eliminated
by local $R$ invariance. The $R$ spectrum of the chiral fermions
could provide an organising principle for the phenomenological
generation structure, with extra insertions needed to produce
couplings of the Higgs boson to the lower generations.\

\section{Modified String Model}

Consider a string presented in cosmic configuration [7]. A feature
of strictly 4 dimensional strings is that they produce a
non-perturbative geometric defect, and also a defect in the
pseudoscalar $\phi$ dual to $B_{\mu \nu}$ around a loop linking
the string worldsheet.

The string tension $T$ and the Regge slope $\alpha '$ for the
(0,1) heterotic string are related by [8]
\begin{equation}\label{T}
    T^{-1}=4\pi \alpha ',
\end{equation}
while $\alpha '$ is related to Newton's constant $G$ by [8]
\begin{equation}\label{alpr}
\alpha ' = \frac{16 \pi G}{g^{2}}.
\end{equation}
The angular deficit around a string in cosmic configuration is
given by [9],
\begin{equation}\label{del}
\delta = 8\pi GT
\end{equation}
using the Einstein field equations.  This gives a non-perturbative
defect around the string site, analogous to the defect around a
massive particle in 2+1 dimensions [10]. We now define a
modification to the properties of a string by applying the
condition for an unbroken $N = 1/2$ supersymmetry around the
string site, using the $R$ gauge connection of local
superconformal symmetry to compensate for the Lorentz connection
defect. This is defined so that the spacetime supersymmetry
components right-moving at the string, i.e. those that couple
locally to the string, are taken to be unbroken by the net effect
of the non-perturbative defects around the string. (This condition
is reminiscent of the partial survival of supersymmetry associated
with BPS saturation, and may indicate a self-duality property.)
This modification would enable supersymmetry breaking to be driven
by mechanisms operating at well below the string tension scale, as
in the following section. For the Lorentz connection around the
string to cancel the $R$ connection on these generators we need

\begin{equation}\label{sup}
    \int_{C}A_{R} = \delta
\end{equation}
where $\delta$ is the deficit angle around the string. The
physical interpretation of eq.(\ref{sup}) is that the strings
incorporate a flux tube of strength $\delta$ for the magnetic
field of $A_{R}$. The field strength $H$ around a cosmic
configuration string may be interpreted as follows. Introducing
cylindrical coordinates $(t,r,\theta,z)$ centred on the string,
$H$ is parallel to the surfaces of constant $\theta$. The coupled
equations for gravity and $H$ are equivalent to an Einstein-Cartan
system with $H$ representing a totally antisymmetric torsion which
Ricci flattens the connection on surfaces of constant $\theta$.
The transverse metric on surfaces with $r,t$ constant remains flat
with azimuthal defect angle $\delta$. (Although a classical static
solution is singular at the string, the quantum string theory is,
as usual, finite.)

As regards $\phi$, a straightforward calculation gives [7]
\begin{equation}\label{C}
 \Delta \equiv \frac{1}{g}  \int_{C}\ast H  = \frac{-g}{8}.
\end{equation}

Suppose now that the dual field $\ast H$ can be described in terms
of the pseudoscalar field $\phi$ as
\begin{equation}\label{HS}
    \ast H = d\phi + C\omega_{R}^{(1)} = d\phi + CA_{R}.
\end{equation}
Here we have extended the global gauge invariance of $\phi$, which
arises since the absolute value of $\phi$ is irrelevant, to a
local symmetry by using the axial vector $U(1)$ gauge field
$A_{R}$, which appears as the axial auxiliary field in
conventional $N=1$ supergravity [4]. By introducing this gauge
compensation of $\phi$ by $A_{R}$ it is now possible for $\phi$ to
be single valued around a string, repairing the defect noted in
[7], giving $C =g \Delta / \delta = -g^2/8\delta$, and we may make
a natural gauge choice $\phi = 0$. This means that local $R$ gauge
symmetry now removes $\phi$ (and thus $B_{\mu \nu}$) from the
physical spectrum.

We now face a problem with the consistency of $R$ symmetry at the
quantum level due to ABJ anomalies [5,6]. Diagrams with one $R$
plus two Yang-Mills or two Lorentz currents, or three $R$ currents
are potentially anomalous. To address this problem we need a
suitable realisation of $R$ symmetry within our model. The $R$
transformation produces a $\gamma_{5}$ `rotation' on the
supersymmetry generator $Q_{\alpha}$ which must be realised on the
right-moving current algebra. One way would be to realise
$\gamma_{5}$ directly, using the Clifford algebra generated by
$DX^{\mu}$. However, all the fermions would then contribute with
the same sign to the potentially anomalous diagrams. In any event,
all the gauge fermions will contribute with the same sign. This
suggests the possibility that the $R$ anomalies are cancelled by
the contribution of the generation and Higgs fermions. Another
possibility would be to invoke a Green-Schwarz cancellation
mechanism [11], but this may not work as $B_{\mu \nu}$ is removed
from the spectrum. As well as the 4 spacetime coordinate
superfields on the string worldsheet, there will be 6 internal
superfields $Y^{m}$. The fermion components $DY^{m}$ generate a
Clifford algebra, which includes a Lie subalgebra $T$ generated by
$[DY^{m},DY^{n}]$. The $R$ charge may be realised as a `handle'
charge within this algebra. This requires that $R$ acts
appropriately on the gravitino $\psi_{\mu}$. The right-moving
factor of $\psi_{mu}$ is a spinor $\psi_{R}$ in the Ramond ground
state, which carries $T$ charges, so we can take $R \in T$. Let
$R_{0}$ be the weight of $\psi_{R}$ in $T$. The most obvious
choice would be to set $R=R_{0}$. However, it seems unlikely that
this gives enough freedom to arrange for all the anomalies to
cancel. However, there will be a $SU(3)$ subalgebra $S$ of $T$
which stabilises $\psi_{R}$, and we may take $R = R_{0} + R_{1}$
with $R_{1} \in S$. It is now possible for the spectrum of $R$ on
the non-gauge fermions to be both biased and dispersed, so that a
complete cancellation of $R$ anomalies may be possible for some
special choices of twisted boundary condition sets for the
strings. The group $U$ generating the twisted boundary conditions
would commute with $R$. Because $R$ lies within the right-moving
current algebra, $R$ will be defined on all states, with chiral
representations (and thus contributions to the anomalies of $R$)
restricted to level zero. Note also that the scalar component of
the conformal compensator $\chi$ has $R = 0$ and so leaves the
breaking of $R$ to $\phi$. A by-product of this construction is
that a realistic generation structure may be organised by the need
for $R$ breaking string-scale insertions of
$\mathrm{exp}(\mathrm{i}\phi)$ for Higgs couplings, leading to
generation indexed coupling matrices of the form
\begin{equation}\label{gen}
    M_{jk} = O(I^{R(j)+R(k)})g
\end{equation}
where $I$ represents an insertion factor. The strings in such a
superconformal model may have a topological significance. Express
the fractional deficit angle as
\begin{equation}\label{del3}
   \frac{\delta}{2\pi} = \frac{1}{\nu}.
\end{equation}
If $\nu = N$, an integer, this suggests an interpretation of the
string worldsheet as a branching site for a $Z_{N}$ bundle, just
as a set of branch points defines a Riemann surface. This
interpretation would be relevant to amplitudes for processes
involving linked strings, possibly selecting the $\Gamma_{N}$
subgroup as the relevant modular invariance group in situations
where only the $N$th power of a Dehn twist would be $Z_{N}$
trivial - otherwise, modular invariance in the presence of a
background string could impose further restrictions on the choice
of twisted boundary conditions. Note that $N$ would determine the
value of the dimensionless bare coupling constant $g$, which can
otherwise take an arbitrary value up to $4\pi$, corresponding to
$\delta = 2\pi$.
\section{Supersymmetry Breaking}
Since we have selected models with $g$ field independent, the
gaugino condensation mechanism for supersymmetry breaking [12,13]
will not work. Consider, however, the supersymmetry variation of
the gravitino field $\psi$ which is given by
\begin{equation}\label{VG}
    \delta \psi = D\epsilon = d\epsilon +
    A_{R}\gamma_{5}\epsilon.
\end{equation}
With the natural gauge choice $\phi = 0$ , we have $A_{R} =
\pi^{-1}D \phi$ and eq.(\ref{VG}) becomes
\begin{equation}\label{VG2}
\delta \psi = d\epsilon + \frac{1}{\pi} \gamma_{5}\epsilon D\phi
\end{equation}
so that the global ($d\epsilon = 0$) variation  of the Goldstino
fermion $\chi$ is given by
\begin{equation}\label{VG3}
\delta \chi = \gamma \cdot \delta \psi =\frac{1}{\pi} \gamma \cdot
(D\phi) \gamma_{5}\epsilon  = \frac{1}{\pi} \delta \zeta
\end{equation}
where $\zeta$ is the superpartner of $\phi$, so that we may
identify $\chi = \pi^{-1}\zeta$, giving a natural candidate for
the Goldstino multiplet in the vacuum state. The Goldstone
supermultiplet and the graviton multiplet together are thus given
by the spin $(1)_{L} \otimes (1 \oplus 1/2)_{R}$ heterotic level
zero fields [14]. The scalar/pseudoscalar part is described by the
complex field $z = \psi + \mathrm{i}\phi$. Supersymmetry breaking
is characterised by the superpotential $\mathcal{G}$. Note that
the gaugino condensation mechanism is ruled out in strictly $D=4$
models by the field independence of $g$.  The normalization of the
kinetic term of $\phi$ through eq.(\ref{KB}), together with the
fact that $\mathcal{G}$ should be independent of $\phi$, which can
be gauged away, gives
\begin{equation}\label{GP}
\mathcal{G} = \frac{1}{2g^{2}}\psi^{2} + \mathcal{G}_{1}
\end{equation}
with $\mathcal{G}_{1}$ independent of $z$. This gives the $z$
space a flat, cylindrical Kahler geometry. $\mathcal{G}$ has a
maximum at $\psi = 0$, with $m_{\psi} =m_{3/2}$.  We may make a
comparison with the situation in conventional $N = 1$ supergravity
where the potential depends on scalar fields through
\begin{equation}\label{V}
   V = \mathrm{e}^{-\mathcal{G }}(\mathrm{Tr}[(\partial \partial ^{\ast} \mathcal{G})^{-1} \partial \mathcal{G}\partial ^{\ast} \mathcal{G}]-3)
\end{equation}
where the (anti-)holomorphic differentials $(\partial^{\ast},
\partial)$ are defined on chiral field space.
For a given gravitino mass $m_{3/2}$ a pure supergravity theory is
obtained in anti-deSitter space with curvature $-3m_{3/2}^2$ and a
flat spacetime is only obtained with the bracing effect of a
nonzero $\partial \mathcal{G}$, with the direction of $\partial
\mathcal{G}$ in field space identifying the Goldstino multiplet.
Our model uses the zero norm tension compensator $\chi$ and the
usual negative norm scalar compensator for conformal symmetry in
conventional $N=1$ supergravity is not present. The Goldstino can
be identified in our model even at a stationary value of
$\mathcal{G}$, where it becomes $\zeta$. Since now $\partial
\mathcal{G} = 0$ at the potential minimum, no cancelling negative
term is required for $\Lambda =0$, and indeed without the negative
norm compensator, eq.(\ref{V}) takes the form
\begin{equation}\label{V2}
 V = \mathrm{e}^{-\mathcal{G }}\mathrm{Tr}[(\partial \partial ^{\ast} \mathcal{G})^{-1} \partial \mathcal{G}\partial ^{\ast} \mathcal{G}]
\end{equation}
and the balance between large terms of opposite sign is no longer
needed. The gravitino mass in the vacuum state will be given by
\begin{equation}\label{MG}
   m_{3/2}^2 = \mathrm{e}^{-\mathcal{G}_{0}}
 \end{equation}
 where $\mathcal{G}_{0}$ is the maximum value of $\mathcal{G}$.
 This gives stable supersymmetry breaking, which would be at a
 realistic scale for a suitable value of $\mathcal{G}_{0}$.
 Such a value might be generated by renormalization group effects.
  The cosmological constant may vanish naturally
 in this model, due to global superconformal symmetry. The global superconformal
invariance of the action under finite rescalings of $g_{\mu \nu}$
rather than of $\hat{g}_{\mu \nu}$ would be used to demonstrate
this. The same brane/flux tube mechanism that cancels the
anomalies of $R$ invariance may also make this symmetry viable at
the quantum level. For this strategy to work, it would be
necessary to apply the principle of global superconformal
invariance to show that the maximum of $\mathcal{G}$ lies on the
surfaces $D^{\alpha} = 0$ for all Yang-Mills generators $\alpha$,
as without $D$ terms the potential eq.(\ref{V2}) would already
give $\Lambda = 0$. \

\

\

\

 \textbf{References}

\

 [1] H.Kawai, D.C.Lewellen and S.H.H.Tye, {\em Nucl. Phys.
B\/}{\bf 288} (1987) 1.

 [2] I.Antoniadis and C.Bachas, {\em Nucl. Phys. B\/}{\bf 298}
(1988) 586.

 [3] I.Antoniadis, J.Ellis, J.S.Hagelin and D.V.Nanopoulos, {\em
Phys. Lett. B\/}{\bf 205} (1988) 459.

 [4] P.van Nieuwenhuizen, {\em Phys. Rep.\/}{\bf 86} (1981) 189.

 [5] S.L.Adler, {\em Phys. Rev.\/}{\bf 177} (1969) 2426.

 [6] J.S.Bell and R.Jackiw, Nuovo Cimento 60A, (1967) 47.

 [7] E.Witten, {\em Phys. Lett. B\/}{\bf 153} (1985) 243.

 [8] D.Gross, J.Harvey, E.Martinec and R.Rohm,
 {\em Nucl. Phys. B\/}{\bf 256} (1985) 253;
 {\bf 267} (1986) 75.

 [9] A.Vilenkin, {\em Phys. Rev. D\/}{\bf 23} (1981) 852.

 [10] G. t'Hooft, {\em Comm. Math. Phys\/}{\bf 117} (1988) 685.

 [11] M.Green and J.H.Schwarz {\em Phys. Lett. B\/}{\bf 149} (1984)
 117.

 [12] S.Ferrara, L.Girardello and H.P.Nilles, {\em Phys. Lett
B\/}{\bf 125} (1983) 457.

 [13] M.Dine, R.Rohm, N.Seiberg and E.Witten, {\em Phys. Lett
B\/}{\bf 156} (1985) 55.

 [14] M.Hewitt, hep-th 0302209.

\end{document}